\documentstyle[12pt,cite,epsf,axodraw]{article}
\title{Noncommutative QED and $\gamma\gamma$ scattering}

\author{Namit Mahajan\thanks{E--mail : nm@ducos.ernet.in, 
nmahajan@physics.du.ac.in} \\
	{\em Department of Physics and Astrophysics,} \\
	 {\em University of Delhi, Delhi-110 007, India.}}

\begin{document}
\maketitle
\begin{abstract}
We study $\gamma\gamma$ scattering in noncommutative QED (NCQED) where
the gauge field has Yang-Mills type coupling, giving new contributions
to the scattering process and making it possible for it to occur at tree
level. The process takes place at one loop level
in the Standard Model (SM) and could be an important signal for physics 
beyond SM. But it is found that the Standard Model contribution
far exceeds the tree level contribution of the noncommutative case. \\ \\
{\bf Keywords}: Noncommutative, gamma-gamma scattering \\ \\
{\bf PACS}: 12.60.-i, 13.40.-f 
\end{abstract}
\begin{section}*{Introduction}
\indent Noncommutativity of a pair of conjugate variables forms
the central theme of quantum mechanics in terms of the Uncertainty Principle.
We are quite familiar with the noncommutativity of rotations in 
ordinary Euclidean space.
The idea of noncommutative (NC) space-time can be traced 
back to the work
of Snyder \cite{snyder}. But more recently, string theory arguments have 
motivated an extensive study of Quantum Field Theory (QFT) on NC spaces
\cite{douglas}. The noncommutativity of space-time is realised by the 
coordinate operators, $x_{\mu}$, satisfying
\begin{equation}
[x_{\mu},x_{\nu}] = \iota\Theta_{\mu\nu}
\end{equation}
with $\Theta_{\mu\nu}
= \theta \epsilon_{\mu\nu}$. $\theta$  is the
noncommutativity parameter with dimensions $(mass)^{-2}$ and $\epsilon
_{\mu\nu}$ is a dimensionless antisymmetric matrix with elements ${\mathcal O}
(1)$. 
The field theories formulated on such spaces are non-local and violate
Lorentz symmetry. The deviation from the standard theory 
manifests as violation of Lorentz invariance.
We can still expect manifest Lorentz invariance for energies satisfying
$E^2\theta << 1$. In the limit $\theta \rightarrow 0$, one expects to 
recover the
standard theory. This is true for the theory at classical level. But at the 
quantum level, the limit $\theta \rightarrow 0$ does not lead to the
commutative theory \cite{armoni}.
 The theory of electrons in a strong magnetic field,
projected to the lowest Landau level, is a classic example of  
NC field theory. \\
\indent
 Various attempts, both theoretical and phenomenological, have been made
to study QFT on NC spaces. The study of perturbative behaviour and
divergence structure \cite{sheraz}, $C$, $P$ and  $T$ properties
and renormalisability \cite{shiekh} of such theories has been undertaken.
It has been shown that quantum theories with time-like noncommuatativities
are not unitary \cite{gomis}. We shall therefore restrict our discussion
to the theories with space-like noncomutativities, although it has been 
shown that light-like noncommutative theories are also free of pathologies
\cite{gomis1}.
To this end, the
coordinate commutator simply reads
\begin{equation}
[x_i,x_j] = \iota\theta\epsilon_{ij}
\end{equation}

\indent There have been attempts to write 
down particle physics
models, in particular SM, on
such NC spaces \cite{connes}.
From a phenomenological point of view, various scattering processes have been
analysed \cite{pheno, hewett} along with the attempts to calculate 
additional contributions to the
precisely measured quantities like anomalous magnetic moment \cite{sh1}
and Lamb shift \cite{sh2} in the noncommutative version of QED. 
\end{section}
\begin{section}*{$\gamma\gamma$ scattering in NCQED}
\indent Consider NCQED i.e. a $U(1)$ noncommutative theory coupled 
to fermions. The noncommutative version of a theory can be written
by replacing the field products by what is called the {\em 'star product'}.
The star ($\ast$) product for any two functions is given by
\begin{equation}
f(x)\ast g(x) = f(x)e^{\frac{\iota}{2}\overleftarrow{\partial_{\alpha}}
                 \Theta^{\alpha\beta}
                     \overrightarrow{\partial_{\beta}}}g(x)
\end{equation}

The NCQED action, using the above line of reasoning, is
\begin{equation}
S_{NCQED} = \int d^Dx \Bigg( -\frac{1}{4g^2}F^{\mu\nu}(x)\ast F_{\mu\nu}(x)
               + \iota\bar{\psi}(x)\gamma^{\mu}\ast D_{\mu}\psi(x)
               - m\bar{\psi}(x)\ast\psi(x)\Bigg)
\end{equation}
where $g$ is the coupling and
\begin{equation}
F_{\mu\nu} = \partial_{\mu}A_{\nu}(x) - \partial_{\nu}A_{\mu}(x)
              + \iota g[A_{\mu}(x),A_{\nu}(x)]_{\ast}
\end{equation}
The covariant derivative is given by
\begin{equation}
D_{\mu}\psi(x) = \partial_{\mu}\psi(x) + \iota gA_{\mu}(x)\ast\psi(x)
\end{equation}
The action is invariant under the noncommutative $U(1)$ transformations
obtained by replacing all the products in the standard transformations
by the corresponding star products.

The noncommutativity is encoded in the star product and from the above
expressions it is quite evident that the field strength, even in the case
of $U(1)$, theory is nonlinear in gauge field and it is precisely 
this nonlinearity
that gives rise to additional vertices for the gauge field. 
It is now a straight forward task to derive the Feynman rules from the 
above action \cite{sh1}, Arfaei et.al \cite{pheno}. 
It is found that apart from generating the 
three and four point vertices for the gauge field self interaction,
each interaction vertex picks up a momentum dependent phase factor,
whose argument typically has the structure $\frac{\iota}{2} p \wedge k$.
The $\wedge$ product, in general, is defined as
\begin{equation}
p \wedge k = p_{\mu} \Theta^{\mu\nu} k_{\nu}
\end{equation}
In the case of theories with only space-like noncommutativities, only the 
space-space elements contribute and using Eq.(2)
it simply reduces to the usual vector cross-product of the two
three momenta i.e.
\begin{equation}
p \wedge k = \vec{p}\times \vec{k}
\end{equation}
\indent The process, $\gamma\gamma \longrightarrow \gamma\gamma$ takes place 
at the one loop level in standard QED as well as SM and thus is quite
suppressed. 
But the presence of Yang-Mills type coupling for the photon field in 
NCQED enables the process to take place at the tree level.
This makes the above process a plausible candidate to look for 
physics beyond SM at the tree level.\\
 
\indent The diagrams contributing to the
 scattering process are
%%%%%%%%%%%%%%%%%%%%%%%%%%%%%%%%%%%%%%%%%%%%%%%%%%%%%%%%   

\vskip 1cm
\begin{center}
 \begin{figure}[htb]
\vspace*{-10ex}
\hspace*{2em}
\begin{tabbing}
\begin{picture}(155,120)(-5.0,-20)
\Photon(0,45)(45,0){3}{6}
	\Text(15,50)[c]{$k_1$}
\ArrowLine(20,40)(30,30)
\Photon(0,-45)(45,0){3}{6}
	\Text(15,-50)[c]{$k_2$}
\ArrowLine(20,-40)(30,-30)
\Photon(45,0)(90,0){3}{5}
\Photon(90,0)(135,45){3}{6}
	\Text(120,50)[c]{$p_1$}
\ArrowLine(105,30)(115,40)
\Photon(90,0)(135,-45){3}{6}
	\Text(120,-50)[c]{$p_2$}
\ArrowLine(105,-30)(115,-40)
\end{picture}
\hskip 1.5cm
\begin{picture}(155,120)(-5.0,-20)
\Photon(0,45)(45,0){3}{6}
	\Text(15,50)[c]{$k_1$}
\ArrowLine(20,40)(30,30)
\Photon(0,-45)(45,0){3}{6}
	\Text(15,-50)[c]{$p_1$}
\ArrowLine(30,-30)(20,-40)
\Photon(45,0)(90,0){3}{5}
\Photon(90,0)(135,45){3}{6}
	\Text(120,50)[c]{$k_2$}
\ArrowLine(115,40)(105,30)
\Photon(90,0)(135,-45){3}{6}
	\Text(120,-50)[c]{$p_2$}
\ArrowLine(105,-30)(115,-40)
\end{picture}
\end{tabbing}
\end{figure}
\end{center}
%%%%%%%%%%%
\vskip 1.5cm
%%%%%%%%%%%%
\begin{center}
 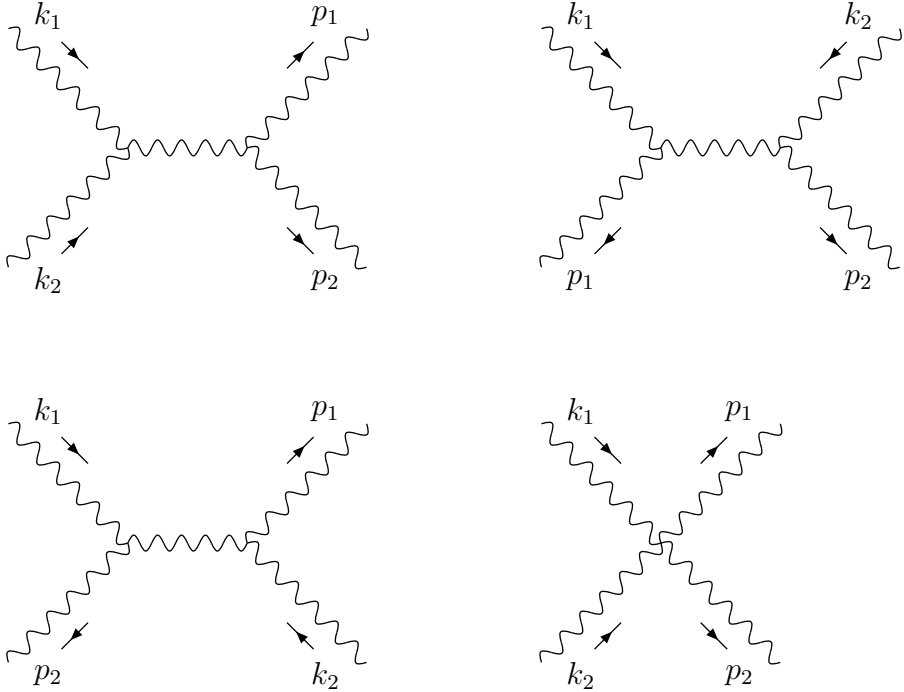
\begin{figure}[htb]
\vspace*{-10ex}
\hspace*{2em}
\begin{tabbing}
\begin{picture}(155,120)(-5.0,-20)
\Photon(0,45)(45,0){3}{6}
	\Text(15,50)[c]{$k_1$}
\ArrowLine(20,40)(30,30)
\Photon(0,-45)(45,0){3}{6}
	\Text(15,-50)[c]{$p_2$}
\ArrowLine(30,-30)(20,-40)
\Photon(45,0)(90,0){3}{5}
\Photon(90,0)(135,45){3}{6}
	\Text(120,50)[c]{$p_1$}
\ArrowLine(105,30)(115,40)
\Photon(90,0)(135,-45){3}{6}
	\Text(120,-50)[c]{$k_2$}
\ArrowLine(115,-40)(105,-30)
\end{picture}
\hskip 1.5cm
\begin{picture}(155,120)(-5.0,-20)
\Photon(0,45)(45,0){3}{6}
	\Text(15,50)[c]{$k_1$}
\ArrowLine(20,40)(30,30)
\Photon(0,-45)(45,0){3}{6}
	\Text(15,-50)[c]{$k_2$}
\ArrowLine(20,-40)(30,-30)
\Photon(45,0)(90,45){3}{6}
	\Text(75,50)[c]{$p_1$}
\ArrowLine(60,30)(70,40)
\Photon(45,0)(90,-45){3}{6}
	\Text(75,-50)[c]{$p_2$}
\ArrowLine(60,-30)(70,-40)
\end{picture}

\end{tabbing}
\vskip 1cm
\caption{$k_1$, $k_2$ denote the incoming photon momenta while $p_1$, 
$p_2$ are the outgoing photon momenta}
\end{figure}
\end{center}
%%%%%%%%%%%%%%%%%%%%%%%%%%%%%%%%%%%%%%%%%%%%%%%%%%%%%%%%%%%%%%%%

Due to the noncommutative nature of the coordinates,
the theory is not Lorentz invariant and the results are frame 
dependent.
In writing down the amplitudes corresponding to each of the 
above diagrams, we assume that $\theta \ll 1$ and make the substitution
$\sin(a\theta) \longrightarrow a\theta$, where $a$ is used to generically 
denote the quantity appearing in the argument of the sine function 
multiplied to $\theta$. \\
\indent Choosing to work in the center of mass frame, we find that
 the s-channel diagram vanishes.
The square of the matrix amplitude reads
\begin{equation}
|{\mathcal{M}}_{NC}|^2 = \left(\frac{e\theta}{16}\right)^4~[100s^4 + 96t^4 +
204st^3 + 360s^2t^2 + 250s^3t]
\end{equation}
and the total (unpolarised) cross section is
\begin{equation}
\sigma_{NC} = (1.5 \times 10^{-3}) \alpha_{em}^2s^3\theta^4
\end{equation}
which for $\sqrt{s}\sim$ TeV and $\theta \leq (10^4~GeV)^{-2}$ 
as in \cite{sh2} 
gives $\sigma_{NC}\sim 10^{-10}$ fb, to
be compared with the SM contribution, $\sigma_{SM}\sim$ fb at the 
same center of mass energy \cite{jikia}. It is found that at low energies
 the fermion 
contribution dominates the SM cross-section while at higher $\sqrt{s}$ ($>$ 
100 GeV), it
is the W contribution that becomes important. The SM contribution gradually
decreases as $\sqrt{s}$ crosses the 500 GeV range. Although, in contrast
to SM, the NC
cross-section increases monotonously with $\sqrt{s}$, it can never catch up
with the SM cross-section for the same energy. 
\end{section}
\begin{section}*{Conclusions}
In this article we have computed the NCQED contribution to the 
$\gamma\gamma$ scattering and found that even though in this case the
process occurs at the tree level as opposed to SM, where
it takes place at the one loop level, the SM contribution far exceeds 
the NC contribution. It is clear that the NCQED contribution will
start showing up only when $\theta$ is much larger than the value
used here.\\ 
\indent The process has been studied in context of NCQED by Hewett 
et.al \cite{hewett}
but the authors argue that inspired by recent theories of extra 
dimensions \cite{nima}, where the effective scale of gravity is 
${\mathcal{O}} \sim$ TeV as opposed to the Planck scale, the scale of
noncommutativity can, too, be chosen to be around TeV. But a more physical
approach would be to use the value of $\theta$, as obtained from
studies like Lamb shift \cite{sh2}, to calculate the 
new contributions. Also the authors have taken into
account time-like noncommutativity that may lead to possible 
non-unitary S-matrix elements.
Even for $\theta~\sim~(TeV)^{-2}$ as taken by the authors,
 the SM contribution is still overwhelmingly large.
Thus with the present day and near future experiments, it doesn't
seem possible to get a signal of NCQED from $\gamma\gamma$ scattering. 
\end{section}
\begin{section}*{Acknowledgements}
The author would like to thank University Grants Commission,
 India for fellowship. 
\end{section} 

\end{document}